\documentclass[aps, pra, superscriptaddress,reprint]{revtex4-1}
\usepackage{amssymb}
\usepackage{hyperref}
\usepackage{amsmath}
\usepackage{graphicx}
\usepackage{bm}
\usepackage[export]{adjustbox}

\begin{document}
\title{Observation of two-beam collective scattering phenomena in a Bose-Einstein condensate}

\author{Ivana Dimitrova}
 \affiliation{Research Laboratory of Electronics, MIT-Harvard Center for Ultracold Atoms,
Department of Physics, Massachusetts Institute of Technology, Cambridge, Massachusetts 02139, USA}
 \author{William Lunden}
  \affiliation{Research Laboratory of Electronics, MIT-Harvard Center for Ultracold Atoms,
Department of Physics, Massachusetts Institute of Technology, Cambridge, Massachusetts 02139, USA}
\author{Jesse Amato-Grill}
\affiliation{Research Laboratory of Electronics, MIT-Harvard Center for Ultracold Atoms,
Department of Physics, Massachusetts Institute of Technology, Cambridge, Massachusetts 02139, USA} 
 \affiliation{Department of Physics, Harvard University, Cambridge, Massachusetts 02138, USA}
 \author{Niklas Jepsen}
  \affiliation{Research Laboratory of Electronics, MIT-Harvard Center for Ultracold Atoms,
Department of Physics, Massachusetts Institute of Technology, Cambridge, Massachusetts 02139, USA}

 \author{Yichao Yu}
 \altaffiliation[Present address: ]{Department of Physics, Harvard University, Cambridge, Massachusetts 02138, USA}
 \affiliation{Research Laboratory of Electronics, MIT-Harvard Center for Ultracold Atoms,
Department of Physics, Massachusetts Institute of Technology, Cambridge, Massachusetts 02139, USA}
 \author{Michael Messer}
 \altaffiliation[Present address: ]{Institute for Quantum Electronics, ETH Zurich, 8093 Zurich, Switzerland}
 \affiliation{Research Laboratory of Electronics, MIT-Harvard Center for Ultracold Atoms,
Department of Physics, Massachusetts Institute of Technology, Cambridge, Massachusetts 02139, USA}
 \author{Thomas Rigaldo}
 \altaffiliation[Present address: ]{Universit\'e Pierre et Marie Curie (Paris 6), 4 place Jussieu, 75 252 Paris Cedex 05, France}
 \affiliation{Research Laboratory of Electronics, MIT-Harvard Center for Ultracold Atoms,
Department of Physics, Massachusetts Institute of Technology, Cambridge, Massachusetts 02139, USA}
 \author{Graciana Puentes}
 \altaffiliation[Present address: ]{(1) Departamento de F\'{\i}sica, FCEyN, UBA, Pabell\'{o}n 1, Ciudad Universitaria, 1428 Buenos Aires, Argentina; (2) Instituto de F\'{\i}sica de Buenos Aires, UBA CONICET, Pabell\'{o}n 1, Ciudad Universitaria, 1428 Buenos Aires, Argentina}
 \affiliation{Research Laboratory of Electronics, MIT-Harvard Center for Ultracold Atoms,
Department of Physics, Massachusetts Institute of Technology, Cambridge, Massachusetts 02139, USA}
 \author{David Weld}
\altaffiliation[Present address: ]{Department of Physics, University of California, Santa Barbara, California 93106, USA} 
 \affiliation{Research Laboratory of Electronics, MIT-Harvard Center for Ultracold Atoms,
Department of Physics, Massachusetts Institute of Technology, Cambridge, Massachusetts 02139, USA}
 \author{Wolfgang Ketterle}
  \affiliation{Research Laboratory of Electronics, MIT-Harvard Center for Ultracold Atoms,
Department of Physics, Massachusetts Institute of Technology, Cambridge, Massachusetts 02139, USA}

\date{\today}

\begin{abstract}
New phenomena of collective light scattering are observed when an elongated Bose-Einstein condensate is pumped by two non-interfering beams counterpropagating along its long axis. In the limit of small Rayleigh scattering rates, the presence of a second pump beam suppresses superradiance, whereas at large Rayleigh scattering rates it lowers the effective threshold power for collective light scattering. In the latter regime, the quench dynamics of the two-beam system are oscillatory, compared to monotonic in the single-beam case. In addition, the dependence on power, detuning, and atom number is explored. The observed features of the two-beam system qualitatively agree with the recent prediction of a supersolid crystalline phase of light and matter at large Rayleigh scattering rates\,\cite{ostermann16}.
\end{abstract}

\maketitle

Collective scattering of light can dramatically enhance single-particle scattering and can lead to qualitatively new phenomena. Since the pioneering work of Dicke \cite{dicke54}, it has been observed in various systems, including thermal atomic and molecular gases\,\cite{gross82, vrehen82, skribanowitz73, macgillivray76, yoshikawa05}, degenerate Bose and Fermi gases\,\cite{inouye99,wang11}, atomic systems in cavities\,\cite{slama07,baumann10}, solid state system\,\cite{scheibner07,cong16} and astrophysical systems\,\cite{rajabi16}. Collective scattering occurs when it is impossible to determine which particle scattered a photon. As a result, correlations develop between the particles which enhance subsequent scattering leading to superradiance. Due to their long coherence time, Bose-Einstein condensates are well suited for studying superradiance\,\cite{inouye99}. Furthermore, the coherence between the atoms, which is responsible for the superradiant scattering, can be directly observed from the momentum distributions and allows a detailed study of superradiance without detection of the emitted light. Different regimes and geometries have been explored by many experimental\,\cite{stenger99, inouye992, yoshikawa05, fallani05, davidson07, hilliard08, li08, deng10, lu11, kampel12, lopes14, muller16} and theoretical studies (see\,\cite{moore99, piovella01a, piovella01b, robb05, zobay05, zobay06} and references therein). In elongated clouds the superradiant gain of the optical mode along the condensate axis, known as the ``end-fire mode,'' is largest, and one can use a single-mode approximation. If the pump laser beam also propagates along this axis, all relevant modes of both the light and the atoms are aligned allowing for a simple one-dimensional description.

\begin{figure}[t!]
\center
\includegraphics[width=0.48 \textwidth]{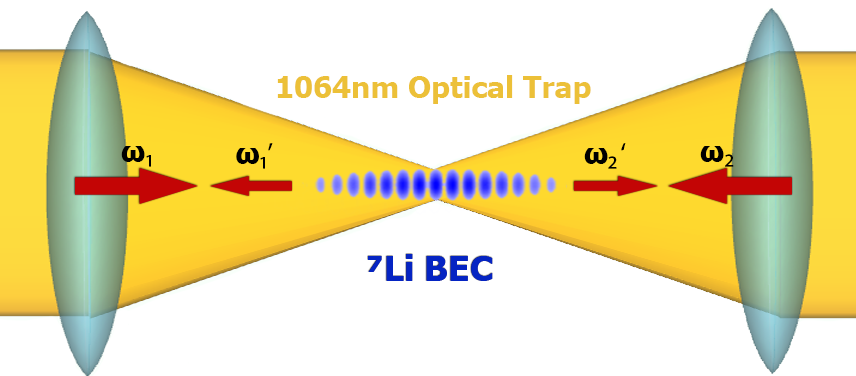}
\caption{Experimental geometry. An elongated Bose-Einstein condensate is exposed to two counterpropagating pump beams detuned by 160\,MHz from each other and by 1 to 20\,GHz from the atomic resonance at 671\,nm. Collective scattering into the end-fire modes creates two backscattered beams and a modulated atomic density distribution. If this distribution is stationary, $\omega_1=\omega_1'$ and $\omega_2=\omega_2'$.}
\label{setup}
\end{figure}

Motivated by recent theoretical work\,\cite{ostermann16}, we have studied the effect of illuminating a Bose-Einstein condensate by two counterpropagating beams at the same intensity. Interference between the two beams can be suppressed by orthogonal polarization or rendered irrelevant by a large frequency offset. Remarkably, we never find a regime in which the two beams act independently, i.e. in which each of them independently induces superradiant scattering into its corresponding direction of propagation. When the Rayleigh scattering rate into the end-fire modes is smaller than the recoil frequency $\omega_R \,{=}\,\hbar k^2/2m$, where $m$ is the atomic mass and $k$ is the wavenumber of the pump light, collective light scattering is suppressed compared to the single-beam case, which is readily explained by bosonic stimulation. By contrast, at high Rayleigh scattering rates, the threshold for collective light scattering is lowered in the presence of a counterpropagating beam. The two regimes are characterized by different microscopic mechanisms. 

At small Rayleigh scattering rates, each scattered photon creates a quasiparticle in the form of a recoiling atom, or a phonon, with momentum equal to the difference between the momenta of the incoming and the scattered photons. Each phonon mode is a \emph{moving} density modulation which is amplified by subsequent collective light scattering. In the geometry of Figure\,\ref{setup}, a single pump beam triggers collective light scattering along the condensate axis resulting in an enhanced number of atoms in the $2 \hbar k$ phonon mode and enhanced backscattered light. An additional counterpropagating beam can create quasiparticles with opposite momentum or annihilate the quasiparticles created by the first beam.

When the Rayleigh scattering rate into the end-fire modes is comparable to or larger than the recoil frequency, two or several photons can be scattered quasi-simultaneously. This creates the possibility of new collective dynamics to emerge, which are no longer governed by individual quasiparticle resonances. The analysis of Ref.\,\cite{ostermann16} shows that above a critical pump power, corresponding to a Rayleigh scattering rate on the order of $\omega_R$, the system develops a type of roton instability towards a periodic modulation and undergoes a phase transition to a crystalline phase. This phase is a \emph{stationary} situation, where an atomic density modulation collectively creates two backscattered beams. Each of them interferes with its corresponding pump beam so that two standing waves are formed. In a self-consistent way, a stationary density modulation is stabilized by the stationary optical lattice potential, which, in contrast to usual optical lattices, consists of two different standing waves which are created and therefore phase-locked by the atoms. In contrast to the single-beam case, in which the backscattered beam is frequency-shifted by 4$\omega_R$ with respect to the incident beam due to the $2 \hbar k$ of recoil momentum imparted to the atoms, in the crystal phase both backscattered beams have the same frequency as their corresponding pump beams because the atomic density distribution is stationary. This establishes a new supersolid form of matter by spontaneous crystallization of light and atoms. For a sudden turn on of the pump beams, there should be an oscillatory behavior around the new equilibrium phase, in contrast to the exponential growth and eventual gain-depletion of single-beam superradiance, as in\,\cite{inouye99}.

In this work, we characterize the regimes of low and high Rayleigh scattering rates by determining the threshold power for collective light scattering and by monitoring the time-evolution of the atomic momentum distribution both for a single beam and for two beams. We observe qualitatively different behaviors, including oscillatory dynamics for two pump beams at high Rayleigh scattering rates. Our observations are consistent with the predictions of the crystal phase using the 1D theoretical model\,\cite{ostermann16}. However, the experimental system is 3D and its lifetime is severely limited by Rayleigh scattering into free space. Therefore, the crystal phase can only form transiently and cannot reach equilibrium.

Experiments are performed with a new $^7$Li machine which produces Bose-Einstein condensates of typically $4\, {\times} \, 10^5$ atoms in 10\,s. Atoms from an effusive oven are laser-cooled with a spin-flip Zeeman slower and $5 \, {\times} \, 10^{9}$ atoms are captured in a 3D Magneto-Optical trap operated on the $D_2$ line. After a compression step, further sub-Doppler cooling is performed using grey molasses on the $D_1$ line, as in\,\cite{grier13}, which reduces the temperature to 90\,$\mu$K. Dark state optical pumping on the $D_1$ line prepares the atoms in the magnetically trappable $|F\,{=}\,2, m_F\,{=}\,2 \rangle$ ground state. They are then transferred to a quadrupole magnetic trap with a repulsive ``plug" optical beam used to inhibit losses from the center of the trap\,\cite{davis95}. During RF evaporation the atomic density is kept at $1\,{\times}\, 10^{13}$ cm$^{-3}$ by gradually opening up the magnetic trap to prevent strong losses due to three-body recombination. The negative scattering length of the $|F\,{=}\,2, m_F\,{=}\,2\rangle$ state prevents the formation of large stable condensates\,\cite{sackett97}. The evaporation is terminated just before degeneracy is reached and the atoms are transferred to a 1064\,nm optical dipole trap. They are spin-flipped to the lowest hyperfine state $|F\,{=}\,1, m_F\,{=}\,1\rangle$ state, which has a Feshbach resonance at 737\,Gauss\,\cite{chin10}. The scattering length is tuned to about 125\,$a_B$, where $a_B$ is the Bohr radius, and the atoms are evaporated to degeneracy.

For the current experiment, a magnetic field is chosen to realize a scattering length of 15\,$a_B$ to avoid strong scattering between atoms in different momentum states. We obtained a cloud of typical size of about $20\, \mu \rm{m}\, {\times}\, 120\, \mu \rm{m}$ by releasing it from a crossed optical dipole trap and letting it expand into a single beam dipole trap. A tunable Ti-Sapph laser generates 671\,nm light detuned by 1 to 20\,GHz from atomic resonance. The two pump beams have $e^{-2}$ radii of $140\, \mu \rm{m}$ and propagate along the long axis of the condensate. Interference between them is suppressed on experimentally relevant timescales by offsetting them in frequency by 160\,MHz using two acousto-optic modulators. This frequency offset is large enough to eliminate Raman coupling between momentum sidebands: $\Delta \omega \, {\gg}\, \omega_R$, and small enough so that the recoil momenta of the two beams are indistinguishable: $\hbar \Delta k \,{ \ll}\, h/L$. Both beams have the same polarization and drive a $\pi$-transition. Rectangular pump pulses are applied, after which the trap is suddenly switched off and momentum distributions are recorded after ballistic expansion. Momentum distributions are then characterized by the number of atoms in the satellite peaks, which are separated from the main cloud by recoil momentum.

\begin{figure}[ht!]
\includegraphics[width=0.48\textwidth]{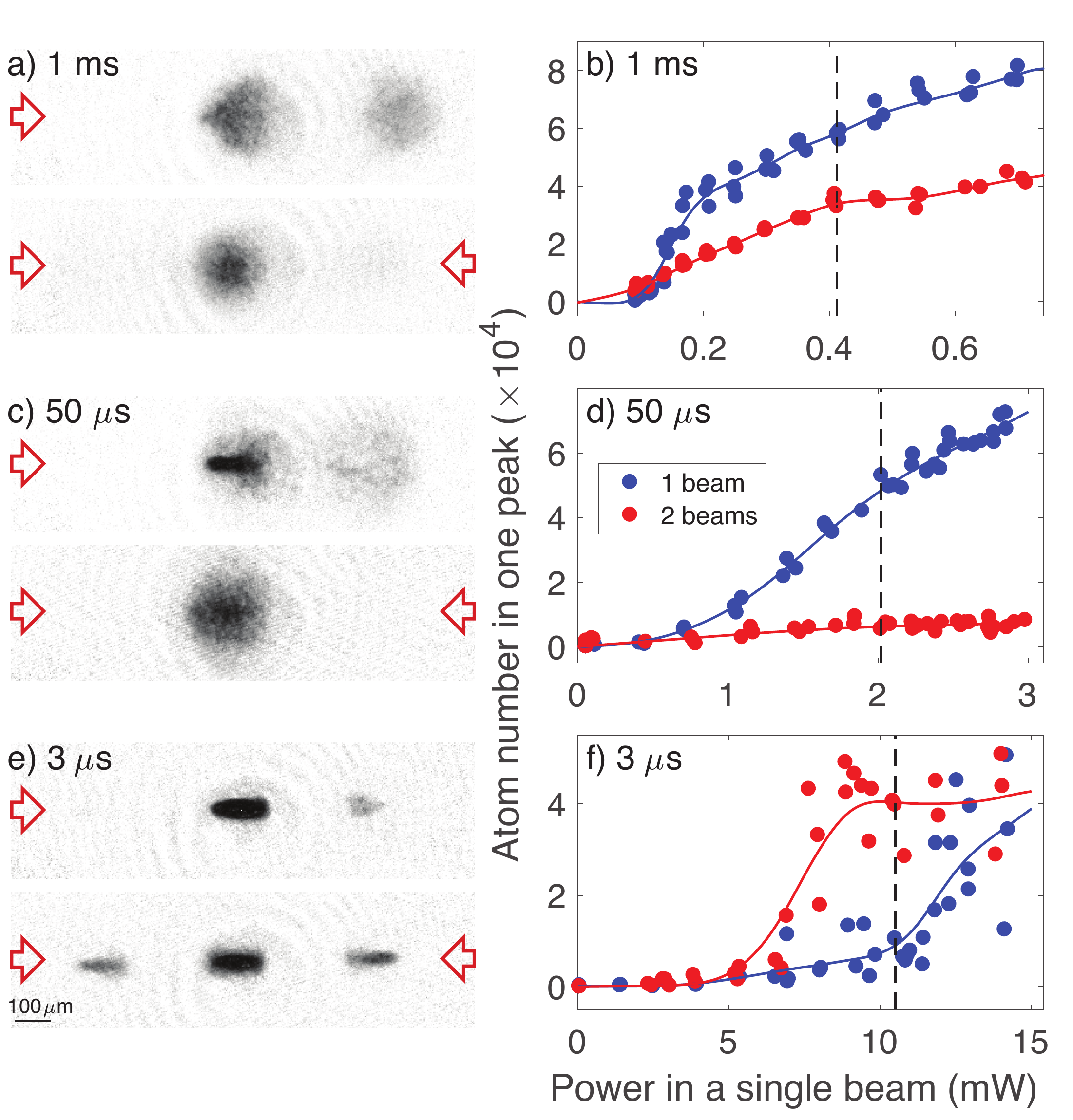}
\caption{Observation of superradiant light scattering in different regimes. Shown are momentum distributions after 2\,ms time of flight for three different pump times and different pump powers at 18.62\,GHz red of the $^7$Li $D_2$ line. For each time, the single-beam case with pump propagating from the left (top picture, triangles) is compared to the case of two balanced beams from opposite sides (bottom, circles). The strength of collective light scattering vs. laser power is characterized by the number of atoms with momentum $2 \hbar k$ (right peak). Solid lines are a guide to the eye. The images are taken at the powers indicated by the dashed vertical lines. }
\label{comparison}
\end{figure}

The onset of collective light scattering is studied by measuring the number of atoms in the $2\hbar k$ peak in time of flight as a function of laser power (Figure\,\ref{comparison}). All three plots on the right side show that there is an effective critical power for the onset of superradiance. For long pump times (Figure\,\ref{comparison}\,a-d), the thresholds are lower, corresponding to small Rayleigh scattering rates. For high pump powers (large Rayleigh scattering rates), the dynamics occur already on short time scales and the critical powers are lower for two beams than for a single beam (Figure\,\ref{comparison}\,e-f). By contrast, at low Rayleigh scattering rates collective light scattering is suppressed in the presence of two beams. This is best seen in Figure\,\ref{comparison}\,c-d: At the same powers, at which the system exhibits strong superradiance with a single pump beam, there are no recoiling atoms visible in the presence of two pump beams.

For Rayleigh scattering rates smaller than the recoil frequency (Figure\,\ref{comparison}\,a-d), a quasiparticle picture can be used to describe the onset of superradiance (for discussion see e.g.\,\cite{inouye99, moore99, robb05, muller16}). Recoiling quasiparticles are created by Rayleigh scattering into the end-fire mode, which occurs at a rate of $R_{ef}=R N_0 f$ where $R$ is the total Rayleigh scattering rate per atom, $N_0$ is the number of atoms in the condensate, and $f$ is the effective solid angle for scattering into the end-fire mode, approximately given by $\lambda^2/D^2$ where $\lambda$ is the wavelength of the scattered light and $D$ is the diameter of the condensate. This rate is enhanced via bosonic stimulation by a factor $N_1+1$ with $N_1$ atoms with recoil momentum $2 \hbar k$ already present. The recoiling atoms are lost from the system at a rate L, either by collisions or because they move out of the condensate volume.

The resulting rate equation describes both the threshold and the initially exponential gain for the case of a sinle pump beam\,\cite{inouye99}:
\begin{equation}
\frac{dN_1}{dt}=R_{ef}(N_1+1) -LN_1=R_{ef}+(RN_0f-L)N_1
\label{SR_rate_eq}
\end{equation}

For weak pump beams and negligible source depletion, one would expect, at least in the perturbative regime, that the addition of a second counterpropagating pump beam would trigger superradiant scattering into the opposite direction. However, due to bosonic stimulation, the rates of scattering into the end-fire modes are proportional to the number of atoms in the initial and the final states. For the case of two counterpropagating beams which can transfer equal but opposite momenta, the stimulated scattering rates, which are responsible for superradiance, cancel in the rate equation
\begin{equation}
\frac{dN_1}{dt}=R_1N_0f(N_1+1)-R_2N_1f(N_0+1)-LN_1 
\end{equation}
if the single-particle scattering rates are equal ($R_1=R_2$). The remaining terms simply reflect spontaneous scattering and loss L as described above. A similar equation can be written for the $- 2 \hbar k$ atoms.

Complete suppression of superradiance for the two-beam case is observed for pump times on the order of tens of recoil times $\omega_R^{-1}\,{=}\,2.5\, \mu$s (e.g. 50\,$\mu$s case in Figure\,\ref{comparison}\,c-d). For even smaller pump powers and therefore longer pump times, as in the 1\,ms data, the suppression is incomplete. This is possibly due to effects of decoherence, or defocusing of the recoiling atoms by atom-atom interactions, which could begin to have an effect after long pump times. For detailed experimental studies of the behavior of the single-beam system at low Rayleigh scattering rates, see\,\cite{inouye99, inouye992, fallani05, li08, hilliard08, deng10, lu11}.

\begin{figure}[h!]
\includegraphics[width=0.48\textwidth]{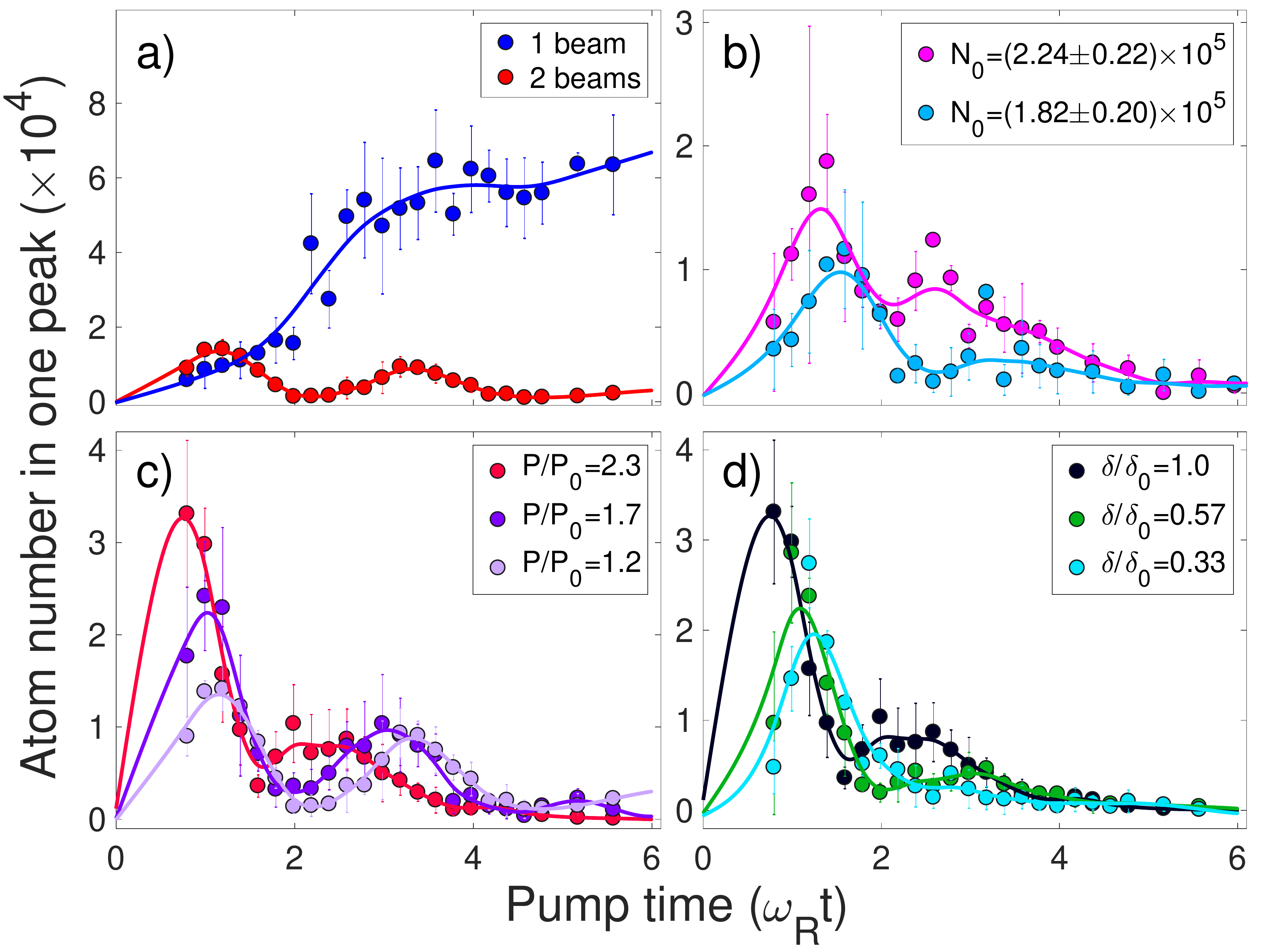}
\caption{Time dynamics of collective light scattering for two beams in the large Rayleigh scattering rate regime. Each beam is at power $P$ which is above the effective threshold power $P_0$ at blue detuning $\delta$ from the $^7$Li $D_2$ line. (a) Comparison of the single-beam pump to the two-beam pump. In both cases, the single-beam power is the same and $P/P_0$\,=\,1.2, where $P_0$\,=\,2.8\,mW at $\delta$\,=\,9.9\,GHz. (b) For different initial condensate numbers ($P/P_0$\,=\,1.8, where $P_0$\,=\,8.6\,mW at $\delta$\,=\,17\,GHz); (c) For different pump powers ($P_0$ and $\delta$ are the same as in (a)); (d) For different detunings but at constant Rayleigh scattering rate, which was measured for a single beam in each case to be 1.7\,$\times 10^4\,\text{s}^{-1}$, which corresponds to $P/P_0 = 2.3$. Here $\delta_0$\,=\,9.9\,GHz. Solid lines are a guide to the eye.}
\label{osc}
\end{figure}

When the Rayleigh scattering rate becomes on the order of the recoil frequency $\omega_R$, the quasiparticle picture can no longer be used. In this regime the system displays the opposite behavior compared to the low scattering rate regime. The presence of the second beam lowers the apparent threshold power for non-zero momentum atoms to appear in time of flight (see Figure\,\ref{comparison}f). In addition, the time dynamics of the system differ qualitatively for the single-beam and the two-beam cases (Figure\,\ref{osc}). When the two beams are suddenly turned on, we observe temporal oscillations of the number of atoms with non-zero momentum. This was predicted by Ref.\,\cite{ostermann16} as oscillations around an equilibrium crystal phase. By contrast, with a single pump beam the number of recoil atoms grows continuously until the Bose-Einstein condensate gets depleted and the gain decreases, as shown in Figure\,\ref{osc}a, in agreement with the predictions of eq.\,(\ref{SR_rate_eq}). The frequency of the oscillations is on the order of the recoil frequency $\omega_R$. The amplitude decays with time due to the loss of atoms to Rayleigh scattering into modes other than the end-fire modes. We observe distinct non-zero momentum peaks in time-of-flight with two beams for about 12\,$\mu$s at an inverse Rayleigh scattering rate of about 30\,$\mu$s. The amplitude and period of the oscillations depend strongly on the initial number of condensate atoms (Figure\,\ref{osc}b). This is characteristic of collective light scattering effects, e.g. in collective spontaneous emission the peak intensity of the emitted light scales with $N^2$, where $N$ is the number of scatterers \cite{vrehen82}. Sample oscillation curves for different initial condensate numbers as well as different loss rates have been numberically calculated in\,\cite{ostermann17} for a 1D two-beam system. The frequency and the amplitude of the oscillations increase with power in the pump beams as shown in Figure\,\ref{osc}c. When the Rayleigh scattering rate is kept constant and the detuning is varied, oscillations with larger amplitude are observed at higher detunings, as evidenced in Figure\,\ref{osc}d. This is consistent with the optical dipole potential rather than Rayleigh scattering governing the dynamics, as in the model used to describe the crystal phase in Ref\,\cite{ostermann16}. A constant Rayleigh scattering rate $R'$ requires the power to scale with detuning $\delta$ as $P '\,{\propto}\, R ' \delta^2$, so that the AC Stark shift and, therefore, the potential depth ${\propto}\, P '/\delta\,{\propto}\, \delta$ \footnote{Oscillations in the intensity of superradiantly emitted light can also occur in a single-beam geometry in the form of collective Rabi oscillations, when the transit time of the light through the cloud is longer than the supperadiant decay time. However, for our experimental conditions, the single-beam case does not display oscillations. Numerical 1D simulations typically predict more ringing than observed in experiments\,\cite{skribanowitz73, kaluzny83, vrehen82, zobay06, muller16}}.

\begin{figure}[h!]\includegraphics[width=0.45\textwidth]{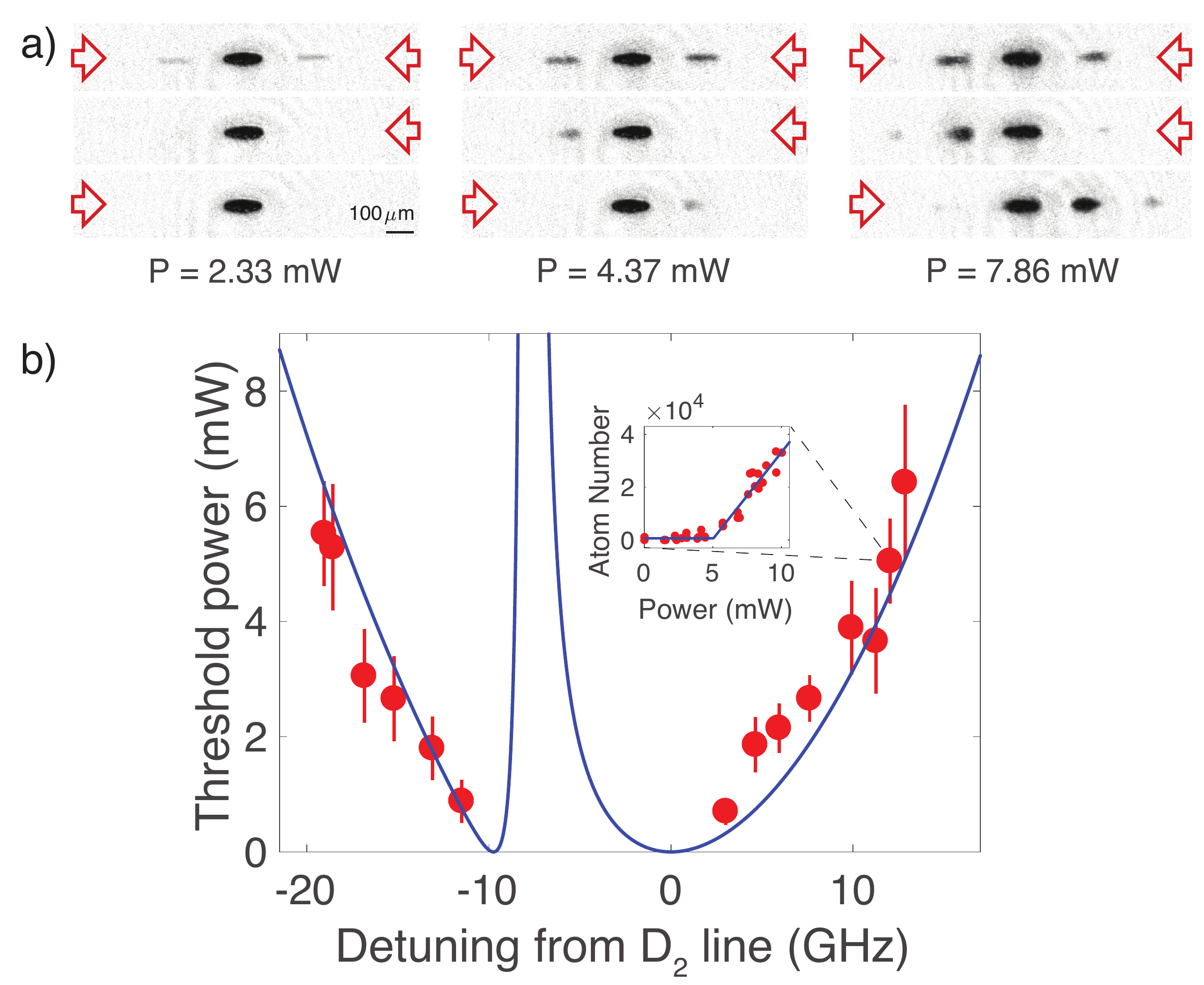}
\caption{(a) Time-of-flight images at three different powers at 3\,$\mu$s pump time, 2\,ms of ballistic expansion, and at 7\,GHz blue of the $D_2$ line. The power noted is the power in each beam. (b) Effective threshold power for the onset of collective light scattering with two pump beams. Each point is obtained from a two-line fit to the number of atoms in the $2 \hbar k$ peak after 3\,$\mu$s pump time as a function of power, as shown in the inset. The error bars are one standard deviation of the threshold fit. The blue line is a quadratic fit for the threshold power versus effective detuning: $P_0=A \tilde{\delta}^2$ with $A$ being a single free parameter.}
\label{discussionfig}
\end{figure}

In addition to the different time dynamics, the single-beam and the two-beam cases also differ in the shape of the observed momentum peaks (Figure\,\ref{discussionfig}a). For the short times of flight used in this experiment, the observed density distributions within each momentum peak still reflect the in-trap density distribution. The shape of the $\pm \,2 \hbar k$ peaks follows closely the elongated shape of the condensate, while the shape of the $2 \hbar k$ peak for a single beam is shorter and more rounded. This can be explained by the inhomogeneous intensity distribution of the backscattered end-fire beam along the condensate. The power of the end-fire mode is largest where the pump beam enters the cloud, increasing the generation of recoiling atoms via Bragg scattering. In contrast, the combined potential of the two stationary optical lattices is predicted to be almost homogeneous along the condensate\,\cite{ostermann16}. At even higher powers (see the rightmost column in Figure\,\ref{discussionfig}a), there is a backward peak of atoms with $-\,2\hbar k$ momentum in the one-beam case. It is the result of the second-order process of re-scattering of a backscattered photon. This is the Kapitza-Dirac regime, commonly associated with pump times smaller than the recoil time\,\cite{schneble03}. We also observe momentum peaks at $4 \hbar k$ due to higher-order superradiance as studied earlier\,\cite{inouye99}. In all cases, the two-beam geometry shows patterns distinctly different from those obtained by adding up the peaks for the two one-beam geometries, indicative of a different mechanism of collective light scattering. 

We have also studied the detuning dependence of the effective threshold power $P_0$ for collective light scattering (Figure\,\ref{discussionfig}b). Due to the fine structure splitting in the excited state (10\,GHz), the effective detuning $\tilde{\delta}$ for a pump beam at frequency $\omega$ is given by $\tilde{\delta}^{-1}\,{=}\,2/(\omega-\omega_{D2})+1/(\omega-\omega_{D1})$, where $\omega_{D1}$ and $\omega_{D2}$ are the frequencies of the $D_1$ and $D_2$ lines and the coefficients reflect the relative strength of the dipole matrix elements for $\pi-$polarized light. The observed thresholds are consistent with $P_0 \propto \tilde{\delta}^2$, i.e. the onset requires a critical Rayleigh scattering rate. This agrees with the threshold power predicted for the crystal phase in\,\cite{ostermann16}. However, due to scatter in the data, the experimental results would also agree almost equally well with a linear fit $P_0 \propto \tilde{\delta}$, i.e. the onset is driven by a critical AC Stark shift. Note that in-between the $D_1$ and $D_2$ lines optical pumping to other hyperfine states limits the gain into the end-fire modes. For the single-beam pump, this leads to Raman superradiance, as observed in\,\citep{schneble04, yoshikawa04}. 

We have so far emphasized the qualitative differences between the one- and two-beam cases. Those cases can be connected by using two beams with imbalanced intensities. In the large Rayleigh scattering limit, we observe that the number of atoms in both satellite peaks initially goes through an oscillation, but eventually the stronger beam wins over: With pump time, one momentum peak grows in number and the other one decreases. Ref.\,\cite{ostermann17} has investigated the phase diagram of the imbalanced system showing that there is a similar instability to self-organization for all values of the beam asymmetry. Further studies are needed to explore this regime. 

In this work, we have observed a self-organization effect of atoms coupled by light. Related effects have been observed with cold atoms in optical cavities (see\,\cite{ritsch13} for review). Superradiant light scattering into a single cavity mode gives rise to only two distinct density modulations, i.e. a checkerboard phase which breaks $\rm{Z_2}$ symmetry. The system studied here is in free space and the density modulation which forms breaks a continuous U(1) symmetry, which constitues a supersolid phase. In fact, a system with two cavities also realizes a supersolid with the spontaneous breaking of a continuous symmetry\,\cite{leonard17}. The present system is conceptually simpler, but the study of the crystal phase is currently limited by the large Rayleigh scattering rates into free space relative to the scattering into the end-fire modes.

In conclusion, we have studied collective light scattering of an elongated Bose-Einstein condensate when pumped with one or two non-interfering beams. In the regime of small Rayleigh scattering rates compared to the recoil frequency, a quasiparticle picture of the scattering explains the suppression of superradiance in the presence of two beams. In the regime of large Rayleigh scattering rates, the behavior of the two-beam system is qualitatively different and consistent with the incipient formation of the predicted crystal phase. The superradiant gain and, as a consequence, the lifetime of the crystal phase could be increased by increasing the atomic density or decreasing the cloud diameter $D$ while keeping the Fresnel number $F \propto D^2/L_0\lambda$ on the order of one, where $L_0$ is the length of the condensate (see\,\cite{gross82} for discussion). This would allow the study of a supersolid formed by collective light scattering, which features a spontaneously chosen phase of the atomic density distribution and the emergent optical lattice. In particular, it would be interesting to confirm the predictions that the backscattered light is not recoil-shifted and that the phase of the density modulation is spontaneously chosen.\\

\begin{acknowledgments}
We would like to thank Helmut Ritsch, Stefan Ostermann, and Francesco Piazza for fruitful discussions and Georgios Siviloglou for help and advice during the initial stages of building the $^7$Li machine. We acknowledge support from the NSF through the Center for Ultracold Atoms and from award 1506369, from ARO-MURI Non-equilibrium Many-body Dynamics (grant W911NF-14-1-0003), from AFOSR-MURI Quantum Phases of Matter (grant FA9550-14-1- 0035), from ONR DURIP (N00014-16-1-3141) and from an ARO seedling grant. J.A.G. acknowledges support by the National Science Foundation Graduate Research Fellowship under Grant No. DGE 1144152. 
\end{acknowledgments}

\bibliography{crystal2bib}

\begin{thebibliography}{42}%
\makeatletter
\providecommand \@ifxundefined [1]{%
 \@ifx{#1\undefined}
}%
\providecommand \@ifnum [1]{%
 \ifnum #1\expandafter \@firstoftwo
 \else \expandafter \@secondoftwo
 \fi
}%
\providecommand \@ifx [1]{%
 \ifx #1\expandafter \@firstoftwo
 \else \expandafter \@secondoftwo
 \fi
}%
\providecommand \natexlab [1]{#1}%
\providecommand \enquote  [1]{``#1''}%
\providecommand \bibnamefont  [1]{#1}%
\providecommand \bibfnamefont [1]{#1}%
\providecommand \citenamefont [1]{#1}%
\providecommand \href@noop [0]{\@secondoftwo}%
\providecommand \href [0]{\begingroup \@sanitize@url \@href}%
\providecommand \@href[1]{\@@startlink{#1}\@@href}%
\providecommand \@@href[1]{\endgroup#1\@@endlink}%
\providecommand \@sanitize@url [0]{\catcode `\\12\catcode `\$12\catcode
  `\&12\catcode `\#12\catcode `\^12\catcode `\_12\catcode `\%12\relax}%
\providecommand \@@startlink[1]{}%
\providecommand \@@endlink[0]{}%
\providecommand \url  [0]{\begingroup\@sanitize@url \@url }%
\providecommand \@url [1]{\endgroup\@href {#1}{\urlprefix }}%
\providecommand \urlprefix  [0]{URL }%
\providecommand \Eprint [0]{\href }%
\providecommand \doibase [0]{http://dx.doi.org/}%
\providecommand \selectlanguage [0]{\@gobble}%
\providecommand \bibinfo  [0]{\@secondoftwo}%
\providecommand \bibfield  [0]{\@secondoftwo}%
\providecommand \translation [1]{[#1]}%
\providecommand \BibitemOpen [0]{}%
\providecommand \bibitemStop [0]{}%
\providecommand \bibitemNoStop [0]{.\EOS\space}%
\providecommand \EOS [0]{\spacefactor3000\relax}%
\providecommand \BibitemShut  [1]{\csname bibitem#1\endcsname}%
\let\auto@bib@innerbib\@empty
\bibitem [{\citenamefont {Ostermann}\ \emph {et~al.}(2016)\citenamefont
  {Ostermann}, \citenamefont {Piazza},\ and\ \citenamefont
  {Ritsch}}]{ostermann16}%
  \BibitemOpen
  \bibfield  {author} {\bibinfo {author} {\bibfnamefont {S.}~\bibnamefont
  {Ostermann}}, \bibinfo {author} {\bibfnamefont {F.}~\bibnamefont {Piazza}}, \
  and\ \bibinfo {author} {\bibfnamefont {H.}~\bibnamefont {Ritsch}},\ }\href
  {\doibase 10.1103/PhysRevX.6.021026} {\bibfield  {journal} {\bibinfo
  {journal} {Phys. Rev. X}\ }\textbf {\bibinfo {volume} {6}},\ \bibinfo {pages}
  {021026} (\bibinfo {year} {2016})}\BibitemShut {NoStop}%
\bibitem [{\citenamefont {Dicke}(1954)}]{dicke54}%
  \BibitemOpen
  \bibfield  {author} {\bibinfo {author} {\bibfnamefont {R.~H.}\ \bibnamefont
  {Dicke}},\ }\href {\doibase 10.1103/PhysRev.93.99} {\bibfield  {journal}
  {\bibinfo  {journal} {Phys. Rev.}\ }\textbf {\bibinfo {volume} {93}},\
  \bibinfo {pages} {99} (\bibinfo {year} {1954})}\BibitemShut {NoStop}%
\bibitem [{\citenamefont {Gross}\ and\ \citenamefont
  {Haroche}(1982)}]{gross82}%
  \BibitemOpen
  \bibfield  {author} {\bibinfo {author} {\bibfnamefont {M.}~\bibnamefont
  {Gross}}\ and\ \bibinfo {author} {\bibfnamefont {S.}~\bibnamefont
  {Haroche}},\ }\href {\doibase http://dx.doi.org/10.1016/0370-1573(82)90102-8}
  {\bibfield  {journal} {\bibinfo  {journal} {Physics Reports}\ }\textbf
  {\bibinfo {volume} {93}},\ \bibinfo {pages} {301 } (\bibinfo {year}
  {1982})}\BibitemShut {NoStop}%
\bibitem [{\citenamefont {Vrehen}\ and\ \citenamefont
  {Gibbs}(1982)}]{vrehen82}%
  \BibitemOpen
  \bibfield  {author} {\bibinfo {author} {\bibfnamefont {Q.~H.~F.}\
  \bibnamefont {Vrehen}}\ and\ \bibinfo {author} {\bibfnamefont {H.~M.}\
  \bibnamefont {Gibbs}},\ }\enquote {\bibinfo {title} {Superfluorescence
  experiments},}\ in\ \href {\doibase 10.1007/978-3-642-81717-5_6} {\emph
  {\bibinfo {booktitle} {Dissipative Systems in Quantum Optics: Resonance
  Fluorescence, Optical Bistability, Superfluorescence}}},\ \bibinfo {editor}
  {edited by\ \bibinfo {editor} {\bibfnamefont {R.}~\bibnamefont {Bonifacio}}}\
  (\bibinfo  {publisher} {Springer Berlin Heidelberg},\ \bibinfo {address}
  {Berlin, Heidelberg},\ \bibinfo {year} {1982})\ pp.\ \bibinfo {pages}
  {111--147}\BibitemShut {NoStop}%
\bibitem [{\citenamefont {Skribanowitz}\ \emph {et~al.}(1973)\citenamefont
  {Skribanowitz}, \citenamefont {Herman}, \citenamefont {MacGillivray},\ and\
  \citenamefont {Feld}}]{skribanowitz73}%
  \BibitemOpen
  \bibfield  {author} {\bibinfo {author} {\bibfnamefont {N.}~\bibnamefont
  {Skribanowitz}}, \bibinfo {author} {\bibfnamefont {I.~P.}\ \bibnamefont
  {Herman}}, \bibinfo {author} {\bibfnamefont {J.~C.}\ \bibnamefont
  {MacGillivray}}, \ and\ \bibinfo {author} {\bibfnamefont {M.~S.}\
  \bibnamefont {Feld}},\ }\href {\doibase 10.1103/PhysRevLett.30.309}
  {\bibfield  {journal} {\bibinfo  {journal} {Phys. Rev. Lett.}\ }\textbf
  {\bibinfo {volume} {30}},\ \bibinfo {pages} {309} (\bibinfo {year}
  {1973})}\BibitemShut {NoStop}%
\bibitem [{\citenamefont {MacGillivray}\ and\ \citenamefont
  {Feld}(1976)}]{macgillivray76}%
  \BibitemOpen
  \bibfield  {author} {\bibinfo {author} {\bibfnamefont {J.~C.}\ \bibnamefont
  {MacGillivray}}\ and\ \bibinfo {author} {\bibfnamefont {M.~S.}\ \bibnamefont
  {Feld}},\ }\href {\doibase 10.1103/PhysRevA.14.1169} {\bibfield  {journal}
  {\bibinfo  {journal} {Phys. Rev. A}\ }\textbf {\bibinfo {volume} {14}},\
  \bibinfo {pages} {1169} (\bibinfo {year} {1976})}\BibitemShut {NoStop}%
\bibitem [{\citenamefont {Yoshikawa}\ \emph {et~al.}(2005)\citenamefont
  {Yoshikawa}, \citenamefont {Torii},\ and\ \citenamefont
  {Kuga}}]{yoshikawa05}%
  \BibitemOpen
  \bibfield  {author} {\bibinfo {author} {\bibfnamefont {Y.}~\bibnamefont
  {Yoshikawa}}, \bibinfo {author} {\bibfnamefont {Y.}~\bibnamefont {Torii}}, \
  and\ \bibinfo {author} {\bibfnamefont {T.}~\bibnamefont {Kuga}},\ }\href
  {\doibase 10.1103/PhysRevLett.94.083602} {\bibfield  {journal} {\bibinfo
  {journal} {Phys. Rev. Lett.}\ }\textbf {\bibinfo {volume} {94}},\ \bibinfo
  {pages} {083602} (\bibinfo {year} {2005})}\BibitemShut {NoStop}%
\bibitem [{\citenamefont {Inouye}\ \emph
  {et~al.}(1999{\natexlab{a}})\citenamefont {Inouye}, \citenamefont
  {Chikkatur}, \citenamefont {Stamper-Kurn}, \citenamefont {Stenger},
  \citenamefont {Pritchard},\ and\ \citenamefont {Ketterle}}]{inouye99}%
  \BibitemOpen
  \bibfield  {author} {\bibinfo {author} {\bibfnamefont {S.}~\bibnamefont
  {Inouye}}, \bibinfo {author} {\bibfnamefont {A.~P.}\ \bibnamefont
  {Chikkatur}}, \bibinfo {author} {\bibfnamefont {D.~M.}\ \bibnamefont
  {Stamper-Kurn}}, \bibinfo {author} {\bibfnamefont {J.}~\bibnamefont
  {Stenger}}, \bibinfo {author} {\bibfnamefont {D.~E.}\ \bibnamefont
  {Pritchard}}, \ and\ \bibinfo {author} {\bibfnamefont {W.}~\bibnamefont
  {Ketterle}},\ }\href {\doibase 10.1126/science.285.5427.571} {\bibfield
  {journal} {\bibinfo  {journal} {Science}\ }\textbf {\bibinfo {volume}
  {285}},\ \bibinfo {pages} {571} (\bibinfo {year}
  {1999}{\natexlab{a}})}\BibitemShut {NoStop}%
\bibitem [{\citenamefont {Wang}\ \emph {et~al.}(2011)\citenamefont {Wang},
  \citenamefont {Deng}, \citenamefont {Hagley}, \citenamefont {Fu},
  \citenamefont {Chai},\ and\ \citenamefont {Zhang}}]{wang11}%
  \BibitemOpen
  \bibfield  {author} {\bibinfo {author} {\bibfnamefont {P.}~\bibnamefont
  {Wang}}, \bibinfo {author} {\bibfnamefont {L.}~\bibnamefont {Deng}}, \bibinfo
  {author} {\bibfnamefont {E.~W.}\ \bibnamefont {Hagley}}, \bibinfo {author}
  {\bibfnamefont {Z.}~\bibnamefont {Fu}}, \bibinfo {author} {\bibfnamefont
  {S.}~\bibnamefont {Chai}}, \ and\ \bibinfo {author} {\bibfnamefont
  {J.}~\bibnamefont {Zhang}},\ }\href {\doibase 10.1103/PhysRevLett.106.210401}
  {\bibfield  {journal} {\bibinfo  {journal} {Phys. Rev. Lett.}\ }\textbf
  {\bibinfo {volume} {106}},\ \bibinfo {pages} {210401} (\bibinfo {year}
  {2011})}\BibitemShut {NoStop}%
\bibitem [{\citenamefont {Slama}\ \emph {et~al.}(2007)\citenamefont {Slama},
  \citenamefont {Bux}, \citenamefont {Krenz}, \citenamefont {Zimmermann},\ and\
  \citenamefont {Courteille}}]{slama07}%
  \BibitemOpen
  \bibfield  {author} {\bibinfo {author} {\bibfnamefont {S.}~\bibnamefont
  {Slama}}, \bibinfo {author} {\bibfnamefont {S.}~\bibnamefont {Bux}}, \bibinfo
  {author} {\bibfnamefont {G.}~\bibnamefont {Krenz}}, \bibinfo {author}
  {\bibfnamefont {C.}~\bibnamefont {Zimmermann}}, \ and\ \bibinfo {author}
  {\bibfnamefont {P.~W.}\ \bibnamefont {Courteille}},\ }\href {\doibase
  10.1103/PhysRevLett.98.053603} {\bibfield  {journal} {\bibinfo  {journal}
  {Phys. Rev. Lett.}\ }\textbf {\bibinfo {volume} {98}},\ \bibinfo {pages}
  {053603} (\bibinfo {year} {2007})}\BibitemShut {NoStop}%
\bibitem [{\citenamefont {Baumann}\ \emph {et~al.}(2010)\citenamefont
  {Baumann}, \citenamefont {Guerlin}, \citenamefont {Brennecke},\ and\
  \citenamefont {Esslinger}}]{baumann10}%
  \BibitemOpen
  \bibfield  {author} {\bibinfo {author} {\bibfnamefont {K.}~\bibnamefont
  {Baumann}}, \bibinfo {author} {\bibfnamefont {C.}~\bibnamefont {Guerlin}},
  \bibinfo {author} {\bibfnamefont {F.}~\bibnamefont {Brennecke}}, \ and\
  \bibinfo {author} {\bibfnamefont {T.}~\bibnamefont {Esslinger}},\ }\href
  {http://dx.doi.org/10.1038/nature09009} {\bibfield  {journal} {\bibinfo
  {journal} {Nature}\ }\textbf {\bibinfo {volume} {464}},\ \bibinfo {pages}
  {1301} (\bibinfo {year} {2010})}\BibitemShut {NoStop}%
\bibitem [{\citenamefont {Scheibner}\ \emph {et~al.}(2007)\citenamefont
  {Scheibner}, \citenamefont {Schmidt}, \citenamefont {Worschech},
  \citenamefont {Forchel}, \citenamefont {Bacher}, \citenamefont {Passow},\
  and\ \citenamefont {Hommel}}]{scheibner07}%
  \BibitemOpen
  \bibfield  {author} {\bibinfo {author} {\bibfnamefont {M.}~\bibnamefont
  {Scheibner}}, \bibinfo {author} {\bibfnamefont {T.}~\bibnamefont {Schmidt}},
  \bibinfo {author} {\bibfnamefont {L.}~\bibnamefont {Worschech}}, \bibinfo
  {author} {\bibfnamefont {A.}~\bibnamefont {Forchel}}, \bibinfo {author}
  {\bibfnamefont {G.}~\bibnamefont {Bacher}}, \bibinfo {author} {\bibfnamefont
  {T.}~\bibnamefont {Passow}}, \ and\ \bibinfo {author} {\bibfnamefont
  {D.}~\bibnamefont {Hommel}},\ }\href {http://dx.doi.org/10.1038/nphys494} {\
  \textbf {\bibinfo {volume} {3}},\ \bibinfo {pages} {106 EP } (\bibinfo {year}
  {2007})}\BibitemShut {NoStop}%
\bibitem [{\citenamefont {Cong}\ \emph {et~al.}(2016)\citenamefont {Cong},
  \citenamefont {Zhang}, \citenamefont {Wang}, \citenamefont {Noe},
  \citenamefont {Belyanin},\ and\ \citenamefont {Kono}}]{cong16}%
  \BibitemOpen
  \bibfield  {author} {\bibinfo {author} {\bibfnamefont {K.}~\bibnamefont
  {Cong}}, \bibinfo {author} {\bibfnamefont {Q.}~\bibnamefont {Zhang}},
  \bibinfo {author} {\bibfnamefont {Y.}~\bibnamefont {Wang}}, \bibinfo {author}
  {\bibfnamefont {G.~T.}\ \bibnamefont {Noe}}, \bibinfo {author} {\bibfnamefont
  {A.}~\bibnamefont {Belyanin}}, \ and\ \bibinfo {author} {\bibfnamefont
  {J.}~\bibnamefont {Kono}},\ }\href {\doibase 10.1364/JOSAB.33.000C80}
  {\bibfield  {journal} {\bibinfo  {journal} {J. Opt. Soc. Am. B}\ }\textbf
  {\bibinfo {volume} {33}},\ \bibinfo {pages} {C80} (\bibinfo {year}
  {2016})}\BibitemShut {NoStop}%
\bibitem [{\citenamefont {Rajabi}\ and\ \citenamefont
  {Houde}(2016)}]{rajabi16}%
  \BibitemOpen
  \bibfield  {author} {\bibinfo {author} {\bibfnamefont {F.}~\bibnamefont
  {Rajabi}}\ and\ \bibinfo {author} {\bibfnamefont {M.}~\bibnamefont {Houde}},\
  }\href {http://stacks.iop.org/0004-637X/826/i=2/a=216} {\bibfield  {journal}
  {\bibinfo  {journal} {The Astrophysical Journal}\ }\textbf {\bibinfo {volume}
  {826}},\ \bibinfo {pages} {216} (\bibinfo {year} {2016})}\BibitemShut
  {NoStop}%
\bibitem [{\citenamefont {Stenger}\ \emph {et~al.}(1999)\citenamefont
  {Stenger}, \citenamefont {Inouye}, \citenamefont {Stamper-Kurn},
  \citenamefont {Chikkatur}, \citenamefont {Pritchard},\ and\ \citenamefont
  {Ketterle}}]{stenger99}%
  \BibitemOpen
  \bibfield  {author} {\bibinfo {author} {\bibfnamefont {J.}~\bibnamefont
  {Stenger}}, \bibinfo {author} {\bibfnamefont {S.}~\bibnamefont {Inouye}},
  \bibinfo {author} {\bibfnamefont {D.}~\bibnamefont {Stamper-Kurn}}, \bibinfo
  {author} {\bibfnamefont {A.}~\bibnamefont {Chikkatur}}, \bibinfo {author}
  {\bibfnamefont {D.}~\bibnamefont {Pritchard}}, \ and\ \bibinfo {author}
  {\bibfnamefont {W.}~\bibnamefont {Ketterle}},\ }\href {\doibase
  10.1007/s003400050818} {\bibfield  {journal} {\bibinfo  {journal} {Applied
  Physics B}\ }\textbf {\bibinfo {volume} {69}},\ \bibinfo {pages} {347}
  (\bibinfo {year} {1999})}\BibitemShut {NoStop}%
\bibitem [{\citenamefont {Inouye}\ \emph
  {et~al.}(1999{\natexlab{b}})\citenamefont {Inouye}, \citenamefont {Pfau},
  \citenamefont {Gupta}, \citenamefont {Chikkatur}, \citenamefont {Gorlitz},
  \citenamefont {Pritchard},\ and\ \citenamefont {Ketterle}}]{inouye992}%
  \BibitemOpen
  \bibfield  {author} {\bibinfo {author} {\bibfnamefont {S.}~\bibnamefont
  {Inouye}}, \bibinfo {author} {\bibfnamefont {T.}~\bibnamefont {Pfau}},
  \bibinfo {author} {\bibfnamefont {S.}~\bibnamefont {Gupta}}, \bibinfo
  {author} {\bibfnamefont {A.~P.}\ \bibnamefont {Chikkatur}}, \bibinfo {author}
  {\bibfnamefont {A.}~\bibnamefont {Gorlitz}}, \bibinfo {author} {\bibfnamefont
  {D.~E.}\ \bibnamefont {Pritchard}}, \ and\ \bibinfo {author} {\bibfnamefont
  {W.}~\bibnamefont {Ketterle}},\ }\href {http://dx.doi.org/10.1038/45194}
  {\bibfield  {journal} {\bibinfo  {journal} {Nature}\ }\textbf {\bibinfo
  {volume} {402}},\ \bibinfo {pages} {641} (\bibinfo {year}
  {1999}{\natexlab{b}})}\BibitemShut {NoStop}%
\bibitem [{\citenamefont {Fallani}\ \emph {et~al.}(2005)\citenamefont
  {Fallani}, \citenamefont {Fort}, \citenamefont {Piovella}, \citenamefont
  {Cola}, \citenamefont {Cataliotti}, \citenamefont {Inguscio},\ and\
  \citenamefont {Bonifacio}}]{fallani05}%
  \BibitemOpen
  \bibfield  {author} {\bibinfo {author} {\bibfnamefont {L.}~\bibnamefont
  {Fallani}}, \bibinfo {author} {\bibfnamefont {C.}~\bibnamefont {Fort}},
  \bibinfo {author} {\bibfnamefont {N.}~\bibnamefont {Piovella}}, \bibinfo
  {author} {\bibfnamefont {M.}~\bibnamefont {Cola}}, \bibinfo {author}
  {\bibfnamefont {F.~S.}\ \bibnamefont {Cataliotti}}, \bibinfo {author}
  {\bibfnamefont {M.}~\bibnamefont {Inguscio}}, \ and\ \bibinfo {author}
  {\bibfnamefont {R.}~\bibnamefont {Bonifacio}},\ }\href {\doibase
  10.1103/PhysRevA.71.033612} {\bibfield  {journal} {\bibinfo  {journal} {Phys.
  Rev. A}\ }\textbf {\bibinfo {volume} {71}},\ \bibinfo {pages} {033612}
  (\bibinfo {year} {2005})}\BibitemShut {NoStop}%
\bibitem [{\citenamefont {Bar-Gill}\ \emph {et~al.}(2007)\citenamefont
  {Bar-Gill}, \citenamefont {Rowen},\ and\ \citenamefont
  {Davidson}}]{davidson07}%
  \BibitemOpen
  \bibfield  {author} {\bibinfo {author} {\bibfnamefont {N.}~\bibnamefont
  {Bar-Gill}}, \bibinfo {author} {\bibfnamefont {E.~E.}\ \bibnamefont {Rowen}},
  \ and\ \bibinfo {author} {\bibfnamefont {N.}~\bibnamefont {Davidson}},\
  }\href {\doibase 10.1103/PhysRevA.76.043603} {\bibfield  {journal} {\bibinfo
  {journal} {Phys. Rev. A}\ }\textbf {\bibinfo {volume} {76}},\ \bibinfo
  {pages} {043603} (\bibinfo {year} {2007})}\BibitemShut {NoStop}%
\bibitem [{\citenamefont {Hilliard}\ \emph {et~al.}(2008)\citenamefont
  {Hilliard}, \citenamefont {Kaminski}, \citenamefont {le~Targat},
  \citenamefont {Olausson}, \citenamefont {Polzik},\ and\ \citenamefont
  {M\"uller}}]{hilliard08}%
  \BibitemOpen
  \bibfield  {author} {\bibinfo {author} {\bibfnamefont {A.}~\bibnamefont
  {Hilliard}}, \bibinfo {author} {\bibfnamefont {F.}~\bibnamefont {Kaminski}},
  \bibinfo {author} {\bibfnamefont {R.}~\bibnamefont {le~Targat}}, \bibinfo
  {author} {\bibfnamefont {C.}~\bibnamefont {Olausson}}, \bibinfo {author}
  {\bibfnamefont {E.~S.}\ \bibnamefont {Polzik}}, \ and\ \bibinfo {author}
  {\bibfnamefont {J.~H.}\ \bibnamefont {M\"uller}},\ }\href {\doibase
  10.1103/PhysRevA.78.051403} {\bibfield  {journal} {\bibinfo  {journal} {Phys.
  Rev. A}\ }\textbf {\bibinfo {volume} {78}},\ \bibinfo {pages} {051403}
  (\bibinfo {year} {2008})}\BibitemShut {NoStop}%
\bibitem [{\citenamefont {Li}\ \emph {et~al.}(2008)\citenamefont {Li},
  \citenamefont {Zhou}, \citenamefont {Yang},\ and\ \citenamefont
  {Chen}}]{li08}%
  \BibitemOpen
  \bibfield  {author} {\bibinfo {author} {\bibfnamefont {J.}~\bibnamefont
  {Li}}, \bibinfo {author} {\bibfnamefont {X.}~\bibnamefont {Zhou}}, \bibinfo
  {author} {\bibfnamefont {F.}~\bibnamefont {Yang}}, \ and\ \bibinfo {author}
  {\bibfnamefont {X.}~\bibnamefont {Chen}},\ }\href {\doibase
  http://dx.doi.org/10.1016/j.physleta.2008.04.049} {\bibfield  {journal}
  {\bibinfo  {journal} {Physics Letters A}\ }\textbf {\bibinfo {volume}
  {372}},\ \bibinfo {pages} {4750 } (\bibinfo {year} {2008})}\BibitemShut
  {NoStop}%
\bibitem [{\citenamefont {Deng}\ \emph {et~al.}(2010)\citenamefont {Deng},
  \citenamefont {Hagley}, \citenamefont {Cao}, \citenamefont {Wang},
  \citenamefont {Luo}, \citenamefont {Wang}, \citenamefont {Payne},
  \citenamefont {Yang}, \citenamefont {Zhou}, \citenamefont {Chen},\ and\
  \citenamefont {Zhan}}]{deng10}%
  \BibitemOpen
  \bibfield  {author} {\bibinfo {author} {\bibfnamefont {L.}~\bibnamefont
  {Deng}}, \bibinfo {author} {\bibfnamefont {E.~W.}\ \bibnamefont {Hagley}},
  \bibinfo {author} {\bibfnamefont {Q.}~\bibnamefont {Cao}}, \bibinfo {author}
  {\bibfnamefont {X.}~\bibnamefont {Wang}}, \bibinfo {author} {\bibfnamefont
  {X.}~\bibnamefont {Luo}}, \bibinfo {author} {\bibfnamefont {R.}~\bibnamefont
  {Wang}}, \bibinfo {author} {\bibfnamefont {M.~G.}\ \bibnamefont {Payne}},
  \bibinfo {author} {\bibfnamefont {F.}~\bibnamefont {Yang}}, \bibinfo {author}
  {\bibfnamefont {X.}~\bibnamefont {Zhou}}, \bibinfo {author} {\bibfnamefont
  {X.}~\bibnamefont {Chen}}, \ and\ \bibinfo {author} {\bibfnamefont
  {M.}~\bibnamefont {Zhan}},\ }\href {\doibase 10.1103/PhysRevLett.105.220404}
  {\bibfield  {journal} {\bibinfo  {journal} {Phys. Rev. Lett.}\ }\textbf
  {\bibinfo {volume} {105}},\ \bibinfo {pages} {220404} (\bibinfo {year}
  {2010})}\BibitemShut {NoStop}%
\bibitem [{\citenamefont {Lu}\ \emph {et~al.}(2011)\citenamefont {Lu},
  \citenamefont {Zhou}, \citenamefont {Vogt}, \citenamefont {Fang},\ and\
  \citenamefont {Chen}}]{lu11}%
  \BibitemOpen
  \bibfield  {author} {\bibinfo {author} {\bibfnamefont {B.}~\bibnamefont
  {Lu}}, \bibinfo {author} {\bibfnamefont {X.}~\bibnamefont {Zhou}}, \bibinfo
  {author} {\bibfnamefont {T.}~\bibnamefont {Vogt}}, \bibinfo {author}
  {\bibfnamefont {Z.}~\bibnamefont {Fang}}, \ and\ \bibinfo {author}
  {\bibfnamefont {X.}~\bibnamefont {Chen}},\ }\href {\doibase
  10.1103/PhysRevA.83.033620} {\bibfield  {journal} {\bibinfo  {journal} {Phys.
  Rev. A}\ }\textbf {\bibinfo {volume} {83}},\ \bibinfo {pages} {033620}
  (\bibinfo {year} {2011})}\BibitemShut {NoStop}%
\bibitem [{\citenamefont {Kampel}\ \emph {et~al.}(2012)\citenamefont {Kampel},
  \citenamefont {Griesmaier}, \citenamefont {Steenstrup}, \citenamefont
  {Kaminski}, \citenamefont {Polzik},\ and\ \citenamefont
  {M\"uller}}]{kampel12}%
  \BibitemOpen
  \bibfield  {author} {\bibinfo {author} {\bibfnamefont {N.~S.}\ \bibnamefont
  {Kampel}}, \bibinfo {author} {\bibfnamefont {A.}~\bibnamefont {Griesmaier}},
  \bibinfo {author} {\bibfnamefont {M.~P.}\ \bibnamefont {Steenstrup}},
  \bibinfo {author} {\bibfnamefont {F.}~\bibnamefont {Kaminski}}, \bibinfo
  {author} {\bibfnamefont {E.~S.}\ \bibnamefont {Polzik}}, \ and\ \bibinfo
  {author} {\bibfnamefont {J.~H.}\ \bibnamefont {M\"uller}},\ }\href {\doibase
  10.1103/PhysRevLett.108.090401} {\bibfield  {journal} {\bibinfo  {journal}
  {Phys. Rev. Lett.}\ }\textbf {\bibinfo {volume} {108}},\ \bibinfo {pages}
  {090401} (\bibinfo {year} {2012})}\BibitemShut {NoStop}%
\bibitem [{\citenamefont {Lopes}\ \emph {et~al.}(2014)\citenamefont {Lopes},
  \citenamefont {Imanaliev}, \citenamefont {Bonneau}, \citenamefont {Ruaudel},
  \citenamefont {Cheneau}, \citenamefont {Boiron},\ and\ \citenamefont
  {Westbrook}}]{lopes14}%
  \BibitemOpen
  \bibfield  {author} {\bibinfo {author} {\bibfnamefont {R.}~\bibnamefont
  {Lopes}}, \bibinfo {author} {\bibfnamefont {A.}~\bibnamefont {Imanaliev}},
  \bibinfo {author} {\bibfnamefont {M.}~\bibnamefont {Bonneau}}, \bibinfo
  {author} {\bibfnamefont {J.}~\bibnamefont {Ruaudel}}, \bibinfo {author}
  {\bibfnamefont {M.}~\bibnamefont {Cheneau}}, \bibinfo {author} {\bibfnamefont
  {D.}~\bibnamefont {Boiron}}, \ and\ \bibinfo {author} {\bibfnamefont {C.~I.}\
  \bibnamefont {Westbrook}},\ }\href {\doibase 10.1103/PhysRevA.90.013615}
  {\bibfield  {journal} {\bibinfo  {journal} {Phys. Rev. A}\ }\textbf {\bibinfo
  {volume} {90}},\ \bibinfo {pages} {013615} (\bibinfo {year}
  {2014})}\BibitemShut {NoStop}%
\bibitem [{\citenamefont {Müller}\ \emph {et~al.}(2016)\citenamefont
  {Müller}, \citenamefont {Witthaut}, \citenamefont {le~Targat},
  \citenamefont {Arlt}, \citenamefont {Polzik},\ and\ \citenamefont
  {Hilliard}}]{muller16}%
  \BibitemOpen
  \bibfield  {author} {\bibinfo {author} {\bibfnamefont {J.~H.}\ \bibnamefont
  {Müller}}, \bibinfo {author} {\bibfnamefont {D.}~\bibnamefont {Witthaut}},
  \bibinfo {author} {\bibfnamefont {R.}~\bibnamefont {le~Targat}}, \bibinfo
  {author} {\bibfnamefont {J.~J.}\ \bibnamefont {Arlt}}, \bibinfo {author}
  {\bibfnamefont {E.~S.}\ \bibnamefont {Polzik}}, \ and\ \bibinfo {author}
  {\bibfnamefont {A.~J.}\ \bibnamefont {Hilliard}},\ }\href
  {http://dx.doi.org/10.1080/09500340.2016.1207815} {\bibfield  {journal}
  {\bibinfo  {journal} {Journal of Modern Optics}\ }\textbf {\bibinfo {volume}
  {63}},\ \bibinfo {pages} {1886} (\bibinfo {year} {2016})}\BibitemShut
  {NoStop}%
\bibitem [{\citenamefont {Moore}\ and\ \citenamefont
  {Meystre}(1999)}]{moore99}%
  \BibitemOpen
  \bibfield  {author} {\bibinfo {author} {\bibfnamefont {M.~G.}\ \bibnamefont
  {Moore}}\ and\ \bibinfo {author} {\bibfnamefont {P.}~\bibnamefont
  {Meystre}},\ }\href {\doibase 10.1103/PhysRevLett.83.5202} {\bibfield
  {journal} {\bibinfo  {journal} {Phys. Rev. Lett.}\ }\textbf {\bibinfo
  {volume} {83}},\ \bibinfo {pages} {5202} (\bibinfo {year}
  {1999})}\BibitemShut {NoStop}%
\bibitem [{\citenamefont {Piovella}\ \emph
  {et~al.}(2001{\natexlab{a}})\citenamefont {Piovella}, \citenamefont
  {Bonifacio}, \citenamefont {McNeil},\ and\ \citenamefont
  {Robb}}]{piovella01a}%
  \BibitemOpen
  \bibfield  {author} {\bibinfo {author} {\bibfnamefont {N.}~\bibnamefont
  {Piovella}}, \bibinfo {author} {\bibfnamefont {R.}~\bibnamefont {Bonifacio}},
  \bibinfo {author} {\bibfnamefont {B.}~\bibnamefont {McNeil}}, \ and\ \bibinfo
  {author} {\bibfnamefont {G.}~\bibnamefont {Robb}},\ }\href {\doibase
  http://dx.doi.org/10.1016/S0030-4018(00)01106-8} {\bibfield  {journal}
  {\bibinfo  {journal} {Optics Communications}\ }\textbf {\bibinfo {volume}
  {187}},\ \bibinfo {pages} {165 } (\bibinfo {year}
  {2001}{\natexlab{a}})}\BibitemShut {NoStop}%
\bibitem [{\citenamefont {Piovella}\ \emph
  {et~al.}(2001{\natexlab{b}})\citenamefont {Piovella}, \citenamefont
  {Gatelli},\ and\ \citenamefont {Bonifacio}}]{piovella01b}%
  \BibitemOpen
  \bibfield  {author} {\bibinfo {author} {\bibfnamefont {N.}~\bibnamefont
  {Piovella}}, \bibinfo {author} {\bibfnamefont {M.}~\bibnamefont {Gatelli}}, \
  and\ \bibinfo {author} {\bibfnamefont {R.}~\bibnamefont {Bonifacio}},\ }\href
  {\doibase http://dx.doi.org/10.1016/S0030-4018(01)01293-7} {\bibfield
  {journal} {\bibinfo  {journal} {Optics Communications}\ }\textbf {\bibinfo
  {volume} {194}},\ \bibinfo {pages} {167 } (\bibinfo {year}
  {2001}{\natexlab{b}})}\BibitemShut {NoStop}%
\bibitem [{\citenamefont {Robb}\ \emph {et~al.}(2005)\citenamefont {Robb},
  \citenamefont {Piovella},\ and\ \citenamefont {Bonifacio}}]{robb05}%
  \BibitemOpen
  \bibfield  {author} {\bibinfo {author} {\bibfnamefont {G.~R.~M.}\
  \bibnamefont {Robb}}, \bibinfo {author} {\bibfnamefont {N.}~\bibnamefont
  {Piovella}}, \ and\ \bibinfo {author} {\bibfnamefont {R.}~\bibnamefont
  {Bonifacio}},\ }\href {http://stacks.iop.org/1464-4266/7/i=4/a=002}
  {\bibfield  {journal} {\bibinfo  {journal} {Journal of Optics B: Quantum and
  Semiclassical Optics}\ }\textbf {\bibinfo {volume} {7}},\ \bibinfo {pages}
  {93} (\bibinfo {year} {2005})}\BibitemShut {NoStop}%
\bibitem [{\citenamefont {Zobay}\ and\ \citenamefont
  {Nikolopoulos}(2005)}]{zobay05}%
  \BibitemOpen
  \bibfield  {author} {\bibinfo {author} {\bibfnamefont {O.}~\bibnamefont
  {Zobay}}\ and\ \bibinfo {author} {\bibfnamefont {G.~M.}\ \bibnamefont
  {Nikolopoulos}},\ }\href {\doibase 10.1103/PhysRevA.72.041604} {\bibfield
  {journal} {\bibinfo  {journal} {Phys. Rev. A}\ }\textbf {\bibinfo {volume}
  {72}},\ \bibinfo {pages} {041604} (\bibinfo {year} {2005})}\BibitemShut
  {NoStop}%
\bibitem [{\citenamefont {Zobay}\ and\ \citenamefont
  {Nikolopoulos}(2006)}]{zobay06}%
  \BibitemOpen
  \bibfield  {author} {\bibinfo {author} {\bibfnamefont {O.}~\bibnamefont
  {Zobay}}\ and\ \bibinfo {author} {\bibfnamefont {G.~M.}\ \bibnamefont
  {Nikolopoulos}},\ }\href {\doibase 10.1103/PhysRevA.73.013620} {\bibfield
  {journal} {\bibinfo  {journal} {Phys. Rev. A}\ }\textbf {\bibinfo {volume}
  {73}},\ \bibinfo {pages} {013620} (\bibinfo {year} {2006})}\BibitemShut
  {NoStop}%
\bibitem [{\citenamefont {Grier}\ \emph {et~al.}(2013)\citenamefont {Grier},
  \citenamefont {Ferrier-Barbut}, \citenamefont {Rem}, \citenamefont
  {Delehaye}, \citenamefont {Khaykovich}, \citenamefont {Chevy},\ and\
  \citenamefont {Salomon}}]{grier13}%
  \BibitemOpen
  \bibfield  {author} {\bibinfo {author} {\bibfnamefont {A.~T.}\ \bibnamefont
  {Grier}}, \bibinfo {author} {\bibfnamefont {I.}~\bibnamefont
  {Ferrier-Barbut}}, \bibinfo {author} {\bibfnamefont {B.~S.}\ \bibnamefont
  {Rem}}, \bibinfo {author} {\bibfnamefont {M.}~\bibnamefont {Delehaye}},
  \bibinfo {author} {\bibfnamefont {L.}~\bibnamefont {Khaykovich}}, \bibinfo
  {author} {\bibfnamefont {F.}~\bibnamefont {Chevy}}, \ and\ \bibinfo {author}
  {\bibfnamefont {C.}~\bibnamefont {Salomon}},\ }\href {\doibase
  10.1103/PhysRevA.87.063411} {\bibfield  {journal} {\bibinfo  {journal} {Phys.
  Rev. A}\ }\textbf {\bibinfo {volume} {87}},\ \bibinfo {pages} {063411}
  (\bibinfo {year} {2013})}\BibitemShut {NoStop}%
\bibitem [{\citenamefont {Davis}\ \emph {et~al.}(1995)\citenamefont {Davis},
  \citenamefont {Mewes}, \citenamefont {Andrews}, \citenamefont {van Druten},
  \citenamefont {Durfee}, \citenamefont {Kurn},\ and\ \citenamefont
  {Ketterle}}]{davis95}%
  \BibitemOpen
  \bibfield  {author} {\bibinfo {author} {\bibfnamefont {K.~B.}\ \bibnamefont
  {Davis}}, \bibinfo {author} {\bibfnamefont {M.~O.}\ \bibnamefont {Mewes}},
  \bibinfo {author} {\bibfnamefont {M.~R.}\ \bibnamefont {Andrews}}, \bibinfo
  {author} {\bibfnamefont {N.~J.}\ \bibnamefont {van Druten}}, \bibinfo
  {author} {\bibfnamefont {D.~S.}\ \bibnamefont {Durfee}}, \bibinfo {author}
  {\bibfnamefont {D.~M.}\ \bibnamefont {Kurn}}, \ and\ \bibinfo {author}
  {\bibfnamefont {W.}~\bibnamefont {Ketterle}},\ }\href {\doibase
  10.1103/PhysRevLett.75.3969} {\bibfield  {journal} {\bibinfo  {journal}
  {Phys. Rev. Lett.}\ }\textbf {\bibinfo {volume} {75}},\ \bibinfo {pages}
  {3969} (\bibinfo {year} {1995})}\BibitemShut {NoStop}%
\bibitem [{\citenamefont {Sackett}\ \emph {et~al.}(1997)\citenamefont
  {Sackett}, \citenamefont {Bradley}, \citenamefont {Welling},\ and\
  \citenamefont {Hulet}}]{sackett97}%
  \BibitemOpen
  \bibfield  {author} {\bibinfo {author} {\bibfnamefont {C.}~\bibnamefont
  {Sackett}}, \bibinfo {author} {\bibfnamefont {C.}~\bibnamefont {Bradley}},
  \bibinfo {author} {\bibfnamefont {M.}~\bibnamefont {Welling}}, \ and\
  \bibinfo {author} {\bibfnamefont {R.}~\bibnamefont {Hulet}},\ }\href
  {\doibase 10.1007/s003400050293} {\bibfield  {journal} {\bibinfo  {journal}
  {Applied Physics B}\ }\textbf {\bibinfo {volume} {65}},\ \bibinfo {pages}
  {433} (\bibinfo {year} {1997})}\BibitemShut {NoStop}%
\bibitem [{\citenamefont {Chin}\ \emph {et~al.}(2010)\citenamefont {Chin},
  \citenamefont {Grimm}, \citenamefont {Julienne},\ and\ \citenamefont
  {Tiesinga}}]{chin10}%
  \BibitemOpen
  \bibfield  {author} {\bibinfo {author} {\bibfnamefont {C.}~\bibnamefont
  {Chin}}, \bibinfo {author} {\bibfnamefont {R.}~\bibnamefont {Grimm}},
  \bibinfo {author} {\bibfnamefont {P.}~\bibnamefont {Julienne}}, \ and\
  \bibinfo {author} {\bibfnamefont {E.}~\bibnamefont {Tiesinga}},\ }\href
  {\doibase 10.1103/RevModPhys.82.1225} {\bibfield  {journal} {\bibinfo
  {journal} {Rev. Mod. Phys.}\ }\textbf {\bibinfo {volume} {82}},\ \bibinfo
  {pages} {1225} (\bibinfo {year} {2010})}\BibitemShut {NoStop}%
\bibitem [{\citenamefont {Ostermann}\ \emph {et~al.}()\citenamefont
  {Ostermann}, \citenamefont {Piazza},\ and\ \citenamefont
  {Ritsch}}]{ostermann17}%
  \BibitemOpen
  \bibfield  {author} {\bibinfo {author} {\bibfnamefont {S.}~\bibnamefont
  {Ostermann}}, \bibinfo {author} {\bibfnamefont {F.}~\bibnamefont {Piazza}}, \
  and\ \bibinfo {author} {\bibfnamefont {H.}~\bibnamefont {Ritsch}},\
  }\href@noop {} {\enquote {\bibinfo {title} {Probing and characterizing the
  growth of a crystal of ultracold bosons and light},}\ }\bibinfo
  {howpublished} {arXiv:1710.05577}\BibitemShut {NoStop}%
\bibitem [{Note1()}]{Note1}%
  \BibitemOpen
  \bibinfo {note} {Oscillations in the intensity of superradiantly emitted
  light can also occur in a single-beam geometry in the form of collective Rabi
  oscillations, when the transit time of the light through the cloud is longer
  than the supperadiant decay time. However, for our experimental conditions,
  the single-beam case does not display oscillations. Numerical 1D simulations
  typically predict more ringing than observed in experiments\protect \tmspace
  +\thinmuskip {.1667em}\cite {skribanowitz73, kaluzny83, vrehen82, zobay06,
  muller16}}\BibitemShut {NoStop}%
\bibitem [{\citenamefont {Schneble}\ \emph {et~al.}(2003)\citenamefont
  {Schneble}, \citenamefont {Torii}, \citenamefont {Boyd}, \citenamefont
  {Streed}, \citenamefont {Pritchard},\ and\ \citenamefont
  {Ketterle}}]{schneble03}%
  \BibitemOpen
  \bibfield  {author} {\bibinfo {author} {\bibfnamefont {D.}~\bibnamefont
  {Schneble}}, \bibinfo {author} {\bibfnamefont {Y.}~\bibnamefont {Torii}},
  \bibinfo {author} {\bibfnamefont {M.}~\bibnamefont {Boyd}}, \bibinfo {author}
  {\bibfnamefont {E.~W.}\ \bibnamefont {Streed}}, \bibinfo {author}
  {\bibfnamefont {D.~E.}\ \bibnamefont {Pritchard}}, \ and\ \bibinfo {author}
  {\bibfnamefont {W.}~\bibnamefont {Ketterle}},\ }\href {\doibase
  10.1126/science.1083171} {\bibfield  {journal} {\bibinfo  {journal}
  {Science}\ }\textbf {\bibinfo {volume} {300}},\ \bibinfo {pages} {475}
  (\bibinfo {year} {2003})}\BibitemShut {NoStop}%
\bibitem [{\citenamefont {Schneble}\ \emph {et~al.}(2004)\citenamefont
  {Schneble}, \citenamefont {Campbell}, \citenamefont {Streed}, \citenamefont
  {Boyd}, \citenamefont {Pritchard},\ and\ \citenamefont
  {Ketterle}}]{schneble04}%
  \BibitemOpen
  \bibfield  {author} {\bibinfo {author} {\bibfnamefont {D.}~\bibnamefont
  {Schneble}}, \bibinfo {author} {\bibfnamefont {G.~K.}\ \bibnamefont
  {Campbell}}, \bibinfo {author} {\bibfnamefont {E.~W.}\ \bibnamefont
  {Streed}}, \bibinfo {author} {\bibfnamefont {M.}~\bibnamefont {Boyd}},
  \bibinfo {author} {\bibfnamefont {D.~E.}\ \bibnamefont {Pritchard}}, \ and\
  \bibinfo {author} {\bibfnamefont {W.}~\bibnamefont {Ketterle}},\ }\href
  {\doibase 10.1103/PhysRevA.69.041601} {\bibfield  {journal} {\bibinfo
  {journal} {Phys. Rev. A}\ }\textbf {\bibinfo {volume} {69}},\ \bibinfo
  {pages} {041601} (\bibinfo {year} {2004})}\BibitemShut {NoStop}%
\bibitem [{\citenamefont {Yoshikawa}\ \emph {et~al.}(2004)\citenamefont
  {Yoshikawa}, \citenamefont {Sugiura}, \citenamefont {Torii},\ and\
  \citenamefont {Kuga}}]{yoshikawa04}%
  \BibitemOpen
  \bibfield  {author} {\bibinfo {author} {\bibfnamefont {Y.}~\bibnamefont
  {Yoshikawa}}, \bibinfo {author} {\bibfnamefont {T.}~\bibnamefont {Sugiura}},
  \bibinfo {author} {\bibfnamefont {Y.}~\bibnamefont {Torii}}, \ and\ \bibinfo
  {author} {\bibfnamefont {T.}~\bibnamefont {Kuga}},\ }\href {\doibase
  10.1103/PhysRevA.69.041603} {\bibfield  {journal} {\bibinfo  {journal} {Phys.
  Rev. A}\ }\textbf {\bibinfo {volume} {69}},\ \bibinfo {pages} {041603}
  (\bibinfo {year} {2004})}\BibitemShut {NoStop}%
\bibitem [{\citenamefont {Ritsch}\ \emph {et~al.}(2013)\citenamefont {Ritsch},
  \citenamefont {Domokos}, \citenamefont {Brennecke},\ and\ \citenamefont
  {Esslinger}}]{ritsch13}%
  \BibitemOpen
  \bibfield  {author} {\bibinfo {author} {\bibfnamefont {H.}~\bibnamefont
  {Ritsch}}, \bibinfo {author} {\bibfnamefont {P.}~\bibnamefont {Domokos}},
  \bibinfo {author} {\bibfnamefont {F.}~\bibnamefont {Brennecke}}, \ and\
  \bibinfo {author} {\bibfnamefont {T.}~\bibnamefont {Esslinger}},\ }\href
  {\doibase 10.1103/RevModPhys.85.553} {\bibfield  {journal} {\bibinfo
  {journal} {Rev. Mod. Phys.}\ }\textbf {\bibinfo {volume} {85}},\ \bibinfo
  {pages} {553} (\bibinfo {year} {2013})}\BibitemShut {NoStop}%
\bibitem [{\citenamefont {L{\'e}onard}\ \emph {et~al.}(2017)\citenamefont
  {L{\'e}onard}, \citenamefont {Morales}, \citenamefont {Zupancic},
  \citenamefont {Esslinger},\ and\ \citenamefont {Donner}}]{leonard17}%
  \BibitemOpen
  \bibfield  {author} {\bibinfo {author} {\bibfnamefont {J.}~\bibnamefont
  {L{\'e}onard}}, \bibinfo {author} {\bibfnamefont {A.}~\bibnamefont
  {Morales}}, \bibinfo {author} {\bibfnamefont {P.}~\bibnamefont {Zupancic}},
  \bibinfo {author} {\bibfnamefont {T.}~\bibnamefont {Esslinger}}, \ and\
  \bibinfo {author} {\bibfnamefont {T.}~\bibnamefont {Donner}},\ }\href
  {http://dx.doi.org/10.1038/nature21067} {\bibfield  {journal} {\bibinfo
  {journal} {Nature}\ }\textbf {\bibinfo {volume} {543}},\ \bibinfo {pages}
  {87} (\bibinfo {year} {2017})}\BibitemShut {NoStop}%
\end{thebibliography}%
\end{document}